\documentclass [10pt] {article}

\usepackage[usenames]{color}

\usepackage [utf8] {inputenc}
\usepackage [T1] {fontenc}

\usepackage {epsfig}

\usepackage {url}
\usepackage {alltt}

\usepackage{amssymb}
\usepackage{amsmath}

\usepackage{listings}

\usepackage [pdftex,
bookmarksopen=true,
backref=false,
colorlinks=true,
linkcolor=blue,
urlcolor=blue,
citecolor=blue,
pdftitle={TechnicalReport},
pdfauthor={F. Boussinot},
pdfsubject={TechnicalReport}] {hyperref}

\newcommand{\code} [1] {{\tt #1}}

\newcommand {\javathreed} {Java3D}

\newcommand {\verlet} {{\it Velocity-Verlet}}

\newcommand {\position} {{\textbf r}}
\newcommand {\speed} {{\textbf v}}
\newcommand {\acceleration} {{\textbf a}}

\newcommand {\email} [1] {{\href {mailto:#1} {#1}}}

\newenvironment{smallsize} [1]
{\vspace{0.2cm}
\scriptsize 
\lstset{
language=Java, 
numbers=right, 
numberstyle=\tiny, 
numbersep=-10pt, 
stepnumber=2, 
}
#1}
{\vspace{0.2cm}}

\newcommand {\mvector} [1] {\overrightarrow {#1}}

\begin {document}
\title { 
\textbf {Reactive Programming \\
of Simulations in Physics} 
}

\newcommand {\fbsignature} {{\sc Fr\'ed\'eric Boussinot}\\
     Mines-ParisTech, Cemef\\
\small  {\email {frederic.boussinot@mines-paristech.fr}}}

\newcommand {\bmsignature} {{\sc Bernard Monasse}\\
    Mines-ParisTech, Cemef\\
\small  {\email {bernard.monasse@mines-paristech.fr}}
}

\newcommand {\jfssignature} {{\sc Jean-Ferdy Susini}\\
   CNAM, Cédric\\
\small  {\email {jean-ferdinand.susini@cnam.fr}}
}

\author {\fbsignature \and \bmsignature \and\jfssignature }



\maketitle
\begin {abstract}
  We consider the Reactive Programming (RP) approach to simulate physical
  systems. The choice of RP is motivated by the fact that RP genuinely
  offers logical parallelism, instantaneously broadcast events, and
  dynamic creation/destruction of parallel components and events. To
  illustrate our approach, we consider the implementation of a
  system of Molecular Dynamics, in the context of Java with
  the Java3D library for 3D visualisation.
\end {abstract}

\paragraph{Keywords.} Concurrency~; parallelism~; reactive programming~;
physics~;  molecular dynamics. 

\section {Introduction}\label {section:introduction}




The essence of programming is to manipulate data structures through procedures \cite{KNUTH}. Objects which group ({\it encapsulate}) data and the procedures ({\it methods}) to use them is at the basis of object-oriented programming. When the decomposition into objects is adequate, object programming is usually considered as leading to more natural and safer programs as it reduces the distance between the programmer's intuition and the program structure. For example, let us consider a physical system made of atoms. In an object-oriented approach, an atom is quite naturally represented by an object containing the atom state (e.g. position, velocity, and mass) and the procedures to transform the atom state (e.g. computing the next position). 

Some systems to simulate physics are coded with objects as LAMMPS, programmed in C++ \cite {LAMMPS}. As more traditional systems programmed in Fortran (e.g. DL\_POLY \cite{DLPOLY}), their central structure basically consists in arrays storing data, and loops to process the array elements in turn. A general issue with such structures is the difficulty to add (and to remove) data, or objects, during the course of a simulation ({\it dynamically}). Some molecular simulation systems are able to deal with dynamic creation of chimical bonds (this is for example the case of LAMMPS, in which the ReaxFF \cite {ReaxFF} potential is implemented), but, to our knowledge, no simulation system is able to deal with the dynamic creation/destruction of general components (e.g. atoms). 

The dynamic creation/destruction of components can be useful to model multi-scale systems \cite {Multiscale}, in which the scale of sub-systems is allowed to change during execution. Let us consider, a molecule simulated at the All-Atom (AA) scale; when the molecule is isolated from the others, il may be possible to switch its scale to Coarse-Grained (CG) and thus to simulate the molecule more efficiently. The inverse scale change may be mandatory when molecules become so close that their interactions must be described at the lowest AA level. In both cases, a change of scale can be seen as the {\it replacement} of the molecule by a new one. For example, the AA to CG change of scale consists in the creation of a new CG molecule, simultaneous with the destruction of the old AA molecule.

In this text, we propose an approach to the implementation of physical simulations in which the dynamic creation or destruction of components can be simply and naturally expressed. The objective is the design of a
Molecular Dynamics (MD) system allowing multi-scale modeling. We use a programming approach, called Reactive Programming (RP) \cite {MDRP}, with a generalized notion of object: an object encapsulates not only its data and methods to manipulate them, but also its {\it behaviour} which can be used to code its interactions with the other objects. In this context, a simulation is structured as an assembly of interacting objects whose behaviours are run in a coordinated way, by an execution machine able to dynamically create and run new objects, and to destroy others.

\subsection* {Structure of the Paper} 

The paper is structured as follows: 
Section \ref {section:rationale} justifies the use of RP for the implementation of physical simulations.
Section \ref {section:reactive-programming} describes RP in general and more precisely the SugarCubes framework.  MD is described in Section \ref {section:molecular-dynamics}. The implementation of the MD simulation system is described in Section \ref {section:md-system}. Related work is considered in Section \ref {section:related-work}, and Section \ref {section:conclusion} concludes the paper, giving several tracks for future work.

\section {Rationale}\label {section:rationale}

When it is not possible, while considering a physical system, to find exact analytical solutions describing its dynamics, the numerical simulation approach becomes mandatory. Numerical simulations are based, first, on a discretization of time and second, on a stepwise {\it resolution method} implementing an integration algorithm.  The possibility to get analytical solutions is basically limited by complexity issues: complex systems cannot usually be analytically solved. This renders numerical simulations an important topics for physics.

\paragraph {Parallelism.} 
At the logical level, it is often the case that the processing of a complex system is facilitated by decomposing it into several sub-systems linked together. The sub-systems can thus be considered as independent components, running and interacting in a coordinated way. Recast in the terminology of informatics, one would say that the sub-systems are run {\it in parallel} in an environment where they share the same notion of time. Here, we are not speaking of real time but of a {\it logical time}, shared by all the sub-systems.  In this approach, the steps of the numerical resolution method used to simulate the system have to be mapped onto the logical time; we shall return on this subject later.  Since we are assuming time to be logical, we should also qualify parallelism as being logical: parallelism is not introduced here to accelerate execution (for example, by using several processors) but to describe a complex system as a set of parallel entities.  The use of real-parallelism (offered by multiprocessor and multicore machines) to accelerate simulations is of course a major issue, but we think that it is basically a matter of {\it optimisation}, while the use of logical parallelism is basically a matter of {\it expressivity}.

\paragraph {Determinism.}
Simulations of physical systems must verify a strong and man\-datory constraint: no energy should be either created or destroyed during the simulation of a closed system (without interaction with the external world). In other words, {\it total energy\footnote{ Total energy is the sum of potential energy and kinetic energy.}must be preserved} during the simulation process. This is a fundamental constraint: it corresponds to the {\it reversibility in time} of the Newton's laws which is at the basis of classical physics. This constraint can be formulated differently: classical physics is {\it deterministic}. In informatics, this has a deep consequence: as physical simulations should be deterministic, one should give the preference to programming languages in which programs are {\it deterministic by construction}. Note however, that parallelism and determinism do not usually go well together; there are very few programming languages which are able to provide both; we shall return on this point later.

\paragraph {Broadcast events.}
In addition to time reversibility, classical physics rests on a second fundamental assumption: forces (gravitation, electrostatic and inter-atomic forces) are instantaneously transmitted\footnote { We place ourselves in a non-relativistic world, where for example simultaneity exists.  }. Instantaneity in our case should accommodate with the presence of the logical time; actually, instantaneity means that action/reaction forces (third Newton's law) should always be exerted during the same instant of the logical time. It is in the nature of classical physical forces to be broadcast everywhere.  In informatics terms, forces correspond to information units that are instantaneously broadcast to all parallel components. This vision of forces actually identifies forces with {\it instantaneously broadcast events}; we shall see that this notion of instantaneously broadcast event exists in the so-called synchronous reactive formalisms, which thus appear to be good candidates to implement physical simulations.

\paragraph {Modularity.}
It may be the case that some components of a system have to be removed from the system because their simulation is no more relevant (imagine for example an object whose distance from the others becomes greater than some fixed threshold, so that its contribution may be considered as negligible). Destruction of components is usually not a big issue in simulations: in order to remove an object from the global system, it may be for example sufficient to stop considering it during the resolution phase. Dually, in some situation, new components may have to be created, for example in chemistry where chemical bonds linking two atoms can appear in the course of a simulation. Dynamic creation is more difficult to deal with than destruction; for example, a new created object must be introduced only at specific steps of the resolution method, in order to avoid inconsistencies.  In informatics terms, both dynamic destruction and dynamic creation of parallel components should be possible during the course of simulations. This possibility is often called {\it modularity}: in a modular system, new components can appear or disappear during execution, without need to change the other components. Note that the notion of broadcast event fits well with modularity as the introduction of a new component listening or producing an event, or the removal of an already existing one, does not affect the communication with the other parallel components.

\paragraph {Hybrid systems.}
There exist {\it hybrid} physical systems which mix continuous and discrete aspects. For example, consider a ball linked by a string to a fixed pivot and turning around the pivot (continuous aspect); one can  then consider the possibility for the string to be broken (discrete aspect). Numerical simulations of hybrid systems are more complex than those of standard systems: in the previous example, the string component has to be removed from the simulation when the destruction of the string occurs, and the simulation of the ball has to switch from a circle to a straight line. In this respect, a hybrid system can be seen as gathering several related systems (for example, the system where the ball is linked to the pivot, and the system where it is free) ; then, the issue becomes to define when and how to switch from one system to another. Note that the simulation of hybrid systems is related to modularity: in the previous example, one can consider that the breaking of the string entails on the one hand the destruction of both the string component and the circular-moving ball, and on the other hand the {\it simultaneous creation} of a straight-moving new ball appearing at the position of the old one.

\paragraph {Resolution method.}
Let us discuss now the relation between logical time and the discretized time of the resolution method. We shall call {\it instant} the basic unit of the logical time; thus, a simulation goes through a first instant, then a second, and so on, until termination\footnote{ Note that actually, there is no conceptual obligation for a simulation to terminate.  }.  The only required property of instants is convergence: all the parallel components terminate at each instant. Execution of instants does not necessarily take the same amount of real time; actually, real time becomes irrelevant for the logic of simulations: the basic simulation time is the logical time, not the real time.  The numerical resolution method works on a time discretized in {\it time-steps} during which forces integration is performed according to Newton's second law. Typically, in simulations of atoms, time-steps have a duration of the order of the femto-second. Several steps of execution may be needed by the resolution algorithm to perform one time-step integration; for example, two steps are needed by the velocity-Verlet integration scheme \cite {Verlet} to integrate forces during one time-step: positions are computed during the first step, and velocities during the second.  Actually, a quite natural scheme associates {\it two} instants with each time-step: during the first instant, each component provides its own information; the global information produced during the first instant is then processed by each component during the second instant. Note that such a two-instant scheme maps quite naturally to the velocity-Verlet method: each step of the resolution method is performed during one instant.

\paragraph {Multi-time aspects.}
The use of the same time-step during the whole simulation is not mandatory : one calls {\it multi-time} a system in which the time-step of the resolution method is allowed to vary during the simulation. 

The change of time-step can be {\it global}, meaning that it concerns all the objects present in the simulation.  This can be helpful for example to get a more accurate simulation when a certain configuration of objects is reached (for example, when objects become confined in a certain volume).

Alternatively, the change of time-step can be {\it local}, i.e. concerning only certain objects, but not all. This means that different components of the same simulation are simultaneously simulated using different time-steps. This could be the case for a system in which {\it diffusion} aspects occur; in such a system, objects in some regions are separated by large distances and evolve freely, simulated with large time-steps, while in other regions, objects are closely interacting and should thus be simulated using smaller time-steps. We will return later on a situation of this kind. We shall call such systems {\it multi-time, multi-step} systems (MTMS systems, for short). A major interest of MTMS is that loosely-coupled objects (with rare interactions) can be simulated during long time periods.




\vspace {0.5cm}

In this text, we consider the Reactive Programming \cite {MDRP} (RP) approach to simulate physical systems. The choice of RP is motivated by the fact that RP genuinely offers logical parallelism, instantaneously broadcast events, and dynamic creation/destruction of parallel components and events. Moreover, we choose a totally deterministic instance of RP, called SugarCubes \cite {fb-jfs:sugar98}, based on the Java programming language: indeed, in SugarCubes, programs are deterministic by construction. To illustrate our approach, we shall consider the implementation of a MTMS simulation system of Molecular Dynamics \cite {MolecularDynamics} (MD) in the context of Java, with the {\javathreed} library \cite {JAVA3D} for 3D visualisation.



\section {Reactive programming} \label {section:reactive-programming}

Reactive Programming \cite{MDRP} (RP) offers a simple framework, with a clear and sound semantics, for expressing logical parallelism. In the RP approach, systems are made of parallel components that share the same {\it instants}. Instants thus define a {\it logical clock}, shared by all components. Parallel components synchronise at each end of instant, and thus execute at the same pace. During instants, components can communicate using {\it instantaneously broadcast events}, which are seen in the same way by all components. There exists several variants of RP, which extend general purpose programming languages (for example, ReactiveC \cite {Boussinotrc91} which extends C, and ReactiveML \cite {RML} which extends the ML language). Among these reactive frameworks is SugarCubes \cite {fb-jfs:sugar98}, which extends Java. In SugarCubes, the parallel operator is very specific: it is totally deterministic, which means that, at each instant, a SugarCubes program has a unique output for each possible input. Actually, in SugarCubes parallelism is implemented in a sequential way.

Due to its ``determinism by construction'', we have choosen to use the SugarCubes framework to implement the MD system; we are going to describe SugarCubes in the rest of the section\footnote{We only describe the notions that are needed for the paper to be self-contained; a complete description of SugarCubes can be found in \cite {fb-jfs:sugar98}.}.

\subsection* {SugarCubes}

The two main SugarCubes classes are \code{Instruction} and \code{Machine}. \code{Instruction} is the class of reactive instructions which are defined with reference to instants, and \code{Machine} is the class of reactive machines which run reactive instructions and define their execution environment.

The main instructions of SugarCubes are the following (their names always start by the prefix \code{SC}):

\begin{itemize}
\item [\bf Void statement: ] \code {SC.nothing} does nothing and immediately terminates.

\item [ \bf Next instant: ] \code {SC.stop} does nothing and suspends the execution of the running thread for the current instant; execution will terminate at the next instant.

\item [ \bf Sequence: ] \code {SC.seq (inst1,inst2)} behaves like \code{inst1} and switches immediately to \code{inst2} as soon as \code{inst1} terminates. 

\item [ \bf Parallelism: ] \code {SC.merge (inst1,inst2)} executes one instant of instructions \code{inst1} and \code{inst2} and terminates if both \code{inst1} and \code{inst2} terminate. Execution always starts by \code{inst1} and switches to \code{inst2} when \code{inst1} either terminates or suspends.

\item [ \bf Cyclic execution: ] \code {SC.loop (inst)} executes cyclically \code{inst}: execution of \code{inst} is immediately restarted as soon as it terminates. One supposes that it is not possible for \code{inst} to terminate at the same instant it is started (otherwise, one would get an {\it instantaneous loop} which would cycle forever during the same instant, preventing thus the reactive machine to detect the end of the current instant). 

\item [ \bf Java code: ] \code{SC.action (jact)} runs the \code{execute} method of the Java action \code{jact} (of type \code{JavaAction}) (this can happen several time, as the action can be in a loop).

\item [ \bf Event generation: ] \code {SC.generate (event,value)} generates \code{event}, with \code{value} as associated value, and immediately terminates. 

\item [ \bf Event waiting: ] \code {SC.await (event)} terminates immediately if \code{event} is present (i.e. it has been previously generated during the current instant), otherwise, execution is suspended waiting either for the generation of \code{event} or for the end of the current instant, detected by the reactive machine.

\item [ \bf Event values: ] \code {SC.callback (event,jcall)} executes the Java callback \code{jcall} (of type \code{JavaCallback}) for each value generated with \code{event} during the current instant. In order not to lose possibly generated values, the execution of the instruction lasts during the whole instant and terminates at the next instant.

\item [ \bf Preemption: ] \code {SC.until (event,inst)} executes \code{inst} and terminates either because \code{inst} terminates, or because \code{event} is present. 

\end {itemize}
The sequence and merge operators are naturally extended to more than two branches; for example \code {SC.seq (i1,i2,i3)} is the sequence of the three instructions \code{i1,i2,i3}.

A reactive machine of the class \code{Machine} runs a program (of type \code{Program}) which is an instruction (initially \code{SC.nothing}). New instructions added to the machine are put in parallel (merge) with the previous program. Addings of new instructions do not occur during the course of an instant, but only at beginnings of instants.

Basically, a machine cyclically runs its program, detects the end of the current instant, that is when all branches of merge instructions are all terminated or suspended, and then goes to the next instant.

Note that the execution of an instruction by a machine during one instant can take several phases: for example, consider the following code, supposing that event \code{e} is not already generated:

\begin{smallsize}
\begin{lstlisting}
SC.merge (
    SC.await e,
    SC.generate (e,null)
)
\end{lstlisting}
\end{smallsize}
Execution switches to the \code{await} instruction (line 2) which is suspended, as \code{e} is not present. Then, execution switches to the \code{generate} instruction (line 3), which produces \code{e} and terminates. The executing machine detects that
execution has to be continued, because one branch of a \code{merge} instruction is suspended, awaiting an event which is present. Thus, the \code{await} instruction is re-executed, and it now terminates, as \code{e} is present. The \code{merge} instruction is also now terminated.

The execution of a program by a machine is totally deterministic: only one trace of execution is possible for a given program. The execution of SugarCubes programs is actually purely sequential: the parallelism presently offered by SugarCubes is a logical one, not a real one; the issue of real parallelism is considered in Sec. \ref{section:conclusion}.

\section {Molecular Dynamics} \label {section:molecular-dynamics}

Numerical simulation at atomic scale predicts system states and properties from a limited number of physical principles, using a numerical resolution method implemented with computers. In Molecular Dynamics (MD) \cite {MolecularDynamics} systems are organic molecules, metallic atoms, or ions. The goal is to determine the temporal evolution of the geometry and energy of atoms.

At the basis of MD is the classical (newtonian) physics, with the fundamental equation:

\begin {equation*}
\mvector {F} = m \mvector {a} 
\end {equation*}

where $\mvector {F}$ is the force applied to a particle of mass $m$ and $\mvector {a}$ is its acceleration (second derivative of the variation of the position, according to time).

A {\it force-field} is composed of several components, called {\it potentials} (of bonds, valence angles, dihedral angles, van der Waals contributions, electrostatic contributions, {\it etc}.) and  is defined by the analytical form of each of these components, and by the parameters caracterizing them. The basic components used to model molecules are the following: 

\begin {itemize}
\item atoms, with 6 degrees of freedom (position and velocity);

\item bonds, which link two atoms belonging to the same molecule; a bond between two atoms $a, b$  tends to maintain constant the distance $ab$.

\item valence angles, which are the angle formed by two adjacent bonds $ba$ et $bc$ in a same molecule; a valence angle tends to maintain constant the angle $\widehat{abc}$. A valence angle is thus concerned by the positions of three atoms.

\item torsion angles (also called {\it dihedral angles}) are defined by four atoms $a, b, c, d$ consecutively linked in the same molecule: $a$ is linked to $b$, $b$ to $c$, and $c$ to $d$; a torsion angle tends to priviledge particular angles between the planes $abc$ and $bcd$. 


\item van der Waals interactions apply between two atoms which either belong to two different molecules, or are not linked by a chain of less than three (or sometimes, four) bonds, if they belong to the same molecule. They are pair potentials.

\end {itemize}

All these potentials depend on the nature of the concerned atoms and are parametrized differently in specific force-fields. Molecular models can also consider electrostatic interactions (Coulomb's law)  which are pair potentials, as van der Waals potentials are; their implementation is close to van der Waals potentials, with a different dependence to distance.

Intra-molecular forces (bonds, valence angles, torsion angles) as well as inter-molecular forces (van der Waals) are conservative: the work between two points does not depend on the path followed by the force between these two points. Thus, forces can be defined as derivatives of scalar fields. From now on, we consider that potentials are scalar fields and we have:

\begin {equation*}
\mvector {F}(\mathbf r) = -\mvector {\nabla} {\cal U} (\mathbf r)
\end {equation*}
where $\mathbf r$ denotes the coordinates of the point on which the force $\mvector {F}(\mathbf r)$ applies, and $\cal U$ is the potential from which the force derives.

The precise definition of the application of forces according to a specific force-field (namely, the OPLS force-field \cite {OPLS}) is described in detail in \cite {2014arXiv1401.1181M}, from which we have taken the overall presentation of MD.

We now describe the rationals for the choice of RP to implement MD.

\subsection*{Reactive programming for molecular dynamics}

The choice of RP, and more specifically of SugarCubes, is motivated by the following reasons:

\begin {itemize}
\item MD systems are composed of separate, interacting components (atoms and molecules). It seems natural to consider that these components execute in parallel. In standard approaches, there is generally a ``big loop'' which considers components in turn (components are placed in an array). This structuration is rather artificial and does not easily support  dynamic changes of the system (for example, additions of new components or removals of old ones, things that one can find in modeling chemical reactions).

\item In MD simulations, time is discrete, and the resolution method which is at the heart of simulations is based on this discrete time. In RP, time is basically discrete, as it is decomposed in instants. Thus, RP makes the discretisation of time which is at the basis of MD very simple. 


\item MD is based on classical (Newtonian) physics which is deterministic. The strict determinism of the parallel operator provided by SugarCubes reflects the fundamental determinism of Newtonian physics. At implementation level, it simplifies debugging (a faulty situation can be simply reproduced). At the physical level, it is mandatory to make simulations reversible in time.

\item In classical physics, interactions are instantaneous which can be quite naturaly expressed using the instantaneously broadcast event notion of RP.

\end {itemize}
In conclusion, the use of RP for MD simulations is motivated by its following characteristics: modularity of logical parallelism, intrinsic discretisation of time due to instants, strict determinism of the parallel operator, instantaneity of events used to code interactions.

Let us now consider the use of RP to implement Molecular Dynamics.

\section {MD system} \label {section:md-system}

A molecular system consists in a set of molecules, each molecule being
made of atoms, bonds, valence angles and torsion angles. In the
approach we propose, the molecule components (atoms, bond, angles) are
{\it programs} that are executed under the supervision of another main
program called a {\it reactive machine}. The reactive machine is in
charge of executing the components in a coordinated way, allowing them
to communicate through {\it events}. Events are broadcast to all the
components run by the reactive machine, that is, all components always
``see'' an event in the same way: either it is present for all
components if it is generated by one of them, or it is absent for all
components if it is not generated during the instant. All events are
reset to absent by the reactive machine at the beginning of each
instant. Values can be associated with event generations. In order to
process the values generated with an event, a component has to wait
during the whole instant, processing the values in turn, as they are
generated.

The reactive machine proceeds in instants: the first instant is
executed, then the second, and so on indefinitely. All components
(atoms, bonds, etc) in the machine are run at each instant and there
is an implicit synchronization (synchronization barrier) of all the
components at the end of each instant. In this way, one is sure that
all component have finished their reaction for the current instant and
have processed all the generated events and all their values before
the next instant can start. Basically, this mode of execution is
synchronous parallelism.

The steps of the resolution method (velocity-Verlet) are identified
with the instants of the reactive machine. The positions of atoms are
computed during one step of the resolution method, and the velocities
during the next step. Actually, at each instant, atoms generate their
position and collect the various forces exerted on them (by bonds,
angles, etc). The new positions are computed from previously collected
information at even\footnote { The choice of even or odd instants is
  arbitrary; the only requirement is actually that instants come in
  pairs.  } instants and the new velocities are computed at odd
instants, following the two-step scheme of the velocity-Verlet
numerical resolution method.

Note that the new positions and velocities are computed by the atom
itself: we say that they are parts of the atom {\it
  behavior}. Strictly speaking, an atom is a structure that
encapsulates data (in particular, position and velocity) together with
a behavior which is a program intended to be run by the reactive
machine in which the atom is added. The good programming practice is
that the atom's behavior is the only component that should access the
atom's data. This discipline entails the absence of time-dependent
errors. As direct access to the atom's data is unwilled, events are
the only means for a component to influence an atom. For example, in
order to apply a force to an atom, a component generates an event
whose value is the force; the atom should wait for the event and
process the generated values; in this way, the atom is able to process
the force applied to it.

The constuction of molecules is a program whose execution adds the
molecule components into the reactive machine.  The main steps to
simulate a molecular system are: 1) define a reactive machine; 2) run
a set of molecules in order to add them in the machine; 3) cyclically
run the machine.

In the rest of this section, we give a brief overview of the various
programs that are used to build a simulation. Note that these are
small pieces of code, that we hope to be natural and easily readable.
This depart from standard MD system descriptions, which are usually
decomposed in procedures whose chaining of calls is poorly specified.
In our approach, the scheduling of the various sub-programs is made
clear and unambiguous.

We shall first describes (generic and specific) atoms, then
intra-molecular components (bonds and angles). We will also consider
inter-molecular interactions. Then, we will explain how molecules are
built from the previous atoms and components.

\subsection {Atoms}
An atom cyclically collects the constraints issues from bonds, valence
angles, and dihedrals, then computes one step of the resolution
method, and finally visualizes itself. This behavior can be preempted
by a kill signal (generated for example when the molecule to which the
atom belongs is destroyed). It is coded by the following SugarCubes
program:

\begin{smallsize}
\begin{lstlisting}
SC.until (killSignal,
    SC.loop (
       SC.seq (
         collection (),
         SC.action (new Resolution (this)),
         SC.action (new Paint3D (this)))))
\end{lstlisting}
\end{smallsize}

A constraint is a force that is added to the atom.  The constraints
are received as values of a specific event associated with the atom
(generation of this event is considered in \ref{subsec:components}).
The collection of constraints is performed by a program which is
returned by the following function \code{collection}:

\begin{smallsize}
\begin{lstlisting}
Program collection ()
   {
      return SC.callback (constraintSignal,
                new CollectConstraints (this));
   }
\end{lstlisting}
\end{smallsize}

\noindent
The \code{CollectConstraints} Java callback is defined by:

\begin{smallsize}
\begin{lstlisting}
public class CollectConstraints implements JavaCallback
{
    final Atom me;
    public void execute (final ReactiveEngine _,final Object args)
	{
	    Vector3d f = (Vector3d)args;
	    Utils.add (me.force,f);
	}
    public CollectConstraints (Atom me)
	{
	    this.me = me;
	}
}
\end{lstlisting}
\end{smallsize}
\code{Vector3d} is the type of 3D vectors. The class \code{Utils}
provides several methods to deal with vectors: \code{vect} creates a
vector between two atoms; \code{normalize} normalizes a vector (same
direction, but unit length); \code {sum} is the vector addition;
\code{perp} is the cross-product of vectors; \code{opposite} defines
the opposite vector; finally, \code{extProd} multiplies a vector by a
scalar.  The addition ``in place'' \code{Utils.add (x,y)} is
equivalent to \code{x = Utils.sum (x,y)}.

We have choosen to define the collection of constraints as a function
(and not to inline its body in the atom behavior) to allow specific
atoms to redefine it (actually, to extend it) for their specific
purpose; this is considered in \ref {subsub:spec-atom}.

Action \code {Resolution} performs the resolution method for the atom;
it is described in \ref {subsub:resolution}. Action \code {Paint3D}
asks for the repainting of the atom; for the sake of simplicity, we do not
consider it here.

\subsubsection {Resolution}\label {subsub:resolution}

The resolution method used is the {\verlet} method \cite{Verlet}. 
Let {\position} be the position (depending of the time) of an atom,
{\speed} its velocity, and {\acceleration} its acceleration. The
{\verlet} method is defined by the following
equations, where $\Delta t$ is a time interval:
\begin{eqnarray*}
&\position (t+\Delta t) = \position (t) + \speed (t)\Delta t + 1/2 \acceleration (t) \Delta t^2\\
&\speed (t+\Delta t) = \speed (t) + 1/2 (\acceleration (t) +
\acceleration (t + \Delta t)) \Delta t
\end{eqnarray*}

Implementation proceeds in two steps:

\begin{enumerate}

\item Compute the velocity at half of the time-step, from previous position
  and acceleration, by: 

\begin{equation} \label{verlet:velocity1}
\speed (t+1/2 \Delta t) = \speed (t) + 1/2
  \acceleration (t) \Delta t
\end{equation} 

Use the result to compute the position at full time-step by: 
\begin{equation} \label{verlet:position}
\position (t+\Delta t) = \position (t) + \speed (t +
  1/2\Delta t)\Delta t
\end{equation}

\item Get acceleration $\acceleration (t+\Delta t)$ from forces applied to the
  atom, and compute velocity at full time-step using the velocity at
  half time-step by: 
\begin{equation} \label{verlet:velocity2}
\speed (t+\Delta t) = \speed
  (t+1/2 \Delta t) + 1/2 \acceleration (t+\Delta t) \Delta t
\end{equation}

\end{enumerate}

In order to allow dynamic introduction of new molecules in the system,
they should only be introduced at instants corresponding to the same
step of the resolution method.  Note that, otherwise, the processing
of LJ forces between atoms belonging to two distinct molecules could
be asymetric, which could introduce fake energy in the system.  One
choses to introduce molecules, and thus atoms, only at even instants.

The {\verlet} resolution is coded by the following class \code{Resolution}:

\begin{smallsize}
\begin{lstlisting}
public class Resolution implements JavaAction
{
  final Atom atom;
  boolean started = false;
  final Vector3d acceleration = new Vector3d ();
  public Resolution (Atom atom)
    {
        this.atom = atom;
    }
  public void execute (final ReactiveEngine _)
    {
      double dt = atom.molecule.context.timeStep;
      boolean evenInstant = (0 == atom.workspace.instant % 2);
      if ( !started && !evenInstant ) return; else started = true;
      if ( dt != 0 ) {
        if ( evenInstant ) step1 (dt); else step2 (dt);
      }
      atom.resetForce ();
    }
  void step1 (double dt)
    {
      Utils.add (atom.velocity,Utils.extProd (0.5*dt,acceleration));
      Utils.add (atom.position,Utils.extProd (dt,atom.velocity));
    }    
  void step2 (double dt)
    {
      Utils.extProd (acceleration,1/atom.mass,atom.force);
      Utils.add (atom.velocity,Utils.extProd (0.5*dt,acceleration));	    
    }
}
\end{lstlisting}
\end{smallsize}
The control of the instant at which atom resolution is started is done
at lines 12-14.

Equation \ref {verlet:velocity1} is coded at line 22. Eq. \ref {verlet:position} is then coded at line 23 (it uses
the previous atom position). The acceleration of the atom is
computed from the force exerted on it (second Newton's law) at line
32. Then, Eq. \ref {verlet:velocity2} is computed at line 28.

Note that, for all atoms, the forces computed during an odd instant
are determined from the positions computed during the previous even
instant. 

\subsubsection {Specific atoms} \label {subsub:spec-atom}
We now consider specific atoms, e.g. carbon atoms, for which we have to deal
with LJ interactions. The \code {collection} function is extended for
this purpose. 
A specific event is defined for each kind of atom, on which atoms
signal their existence. In this way, an atom can collect all the signaling
events and compute the forces induced by the LJ interactions with the
other atoms.
The collection function of a carbon atom is for example defined by:

\begin{smallsize}
\begin{lstlisting}
Program collection ()
   {
      return SC.seq ( 
          SC.generate (CSignal,this),
          SC.merge (
              super.collection (),
              collectLJ (CSignal,new LJPotential (ljC_C)),
	      collectLJ (HSignal,new LJPotential (ljC_H)),	    
              collectLJ (OSignal,new LJPotential (ljC_O))));
   }
\end{lstlisting}
\end{smallsize}
Note that this definition actually extends the previous \code{collection}
method of standard atom; this method continues to be called
(\code{super.collection ()}) but is now put in parallel with the
specific treatments of LJ interactions. The collection of the
interactions corresponding to a specific kind of
atoms is coded by:

\begin{smallsize}
\begin{lstlisting}
Program collectLJ (Identifier signal,Potential potential)
   {
      return SC.callback (signal,
         new CollectInteractions (potential,this));
   }
\end{lstlisting}
\end{smallsize}
The \code {CollectInteractions} callback applies the \code
{computeForce} method of the potential parameter to all the atoms
(except itself) which signal their presence through the 
parameter signal, and adds the obtained force to the previously
collected forces.

\subsection {Intra-molecular forces}\label {subsec:components}
We now consider the way intra-molecular forces are produced and
applied to atoms. The application of forces to atoms from a potential is defined in
\cite {2014arXiv1401.1181M}. Here, we shall only consider bonds which
are the simplest components. The treatment of the others components (valence and
torsion angles) is very similar.

\begin{figure} [!htb]
\begin {center}
\includegraphics[width=8cm] {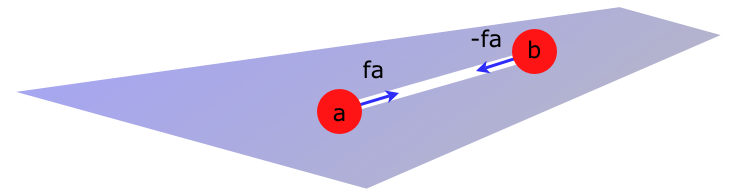}
\end {center}
\caption {Attractive bond between two linked atoms}
\label {figure:liaison}
\end{figure}

A {\it harmonic bond potential} is a scalar field $\cal
U$ which defines the potential energy of two atoms placed at distance
$r$ as:
\begin{equation} \label {bond:potential}
{\cal U}(r) = k (r-r_0)^2
\end{equation}
where $k$ is the strength of the bond and $r_0$ is the equilibrium
distance (the distance at which the force between the two atoms is
null). We thus have:
\begin{equation} \label {bond:force}
{\frac {\partial {\cal U} (r)} {\partial {r}}} = 2k (r - r_0)
\end{equation}

Bonds are coded by the class \code{HarmonicBond} which has the
following behavior:

\begin{smallsize}
\begin{lstlisting}
SC.loop (
   SC.seq (
      SC.action (new ControlLength ()),
      SC.generate (first.constraintSignal,fa),
      SC.generate (second.constraintSignal,fb),
      SC.action (new Paint3D (this)),
      SC.stop () )  )
\end{lstlisting}
\end{smallsize}
The \code{ControlLength} action is called at each instant, to
determine the force to be applied to the two atoms linked by the bond.
The application of forces is realized through the
\code{constraintSignal} of the two atoms. The applied forces are the
values generated with these events. Note the presence of the
\code{stop} statement to avoid an instantaneous loop (which would
produce a warning message at each instant). The \code{ControlLength}
action sets the \code{force} field of class \code{HarmonicBond} and is
defined by:

\begin{smallsize}
\begin{lstlisting}
public class ControlLength implements JavaAction
{
    public void execute (final ReactiveEngine _)
       {
           double dist = Utils.distance (a,b);
           double diff = dist - length;
           energy = strength * diff * diff;
           dUdr = 2.0 * strength * diff;
           Vector3d v12 = Utils.vect (a,b);
           v12.normalize ();
           Utils.extProd (fa,dUdr,v12);
           Utils.extProd (fb,-dUdr,v12);
       }
}
\end{lstlisting}
\end{smallsize}
Eq. \ref {bond:potential} is coded at line 7, and Eq. \ref
{bond:force} at line 8. The force to be applied to the first atom is
computed at line 11 (the force to be applied to the second is the opposite).

\subsection {Molecules}
\begin{figure} [!htb]
\begin {center}
\includegraphics[width=2.5cm] {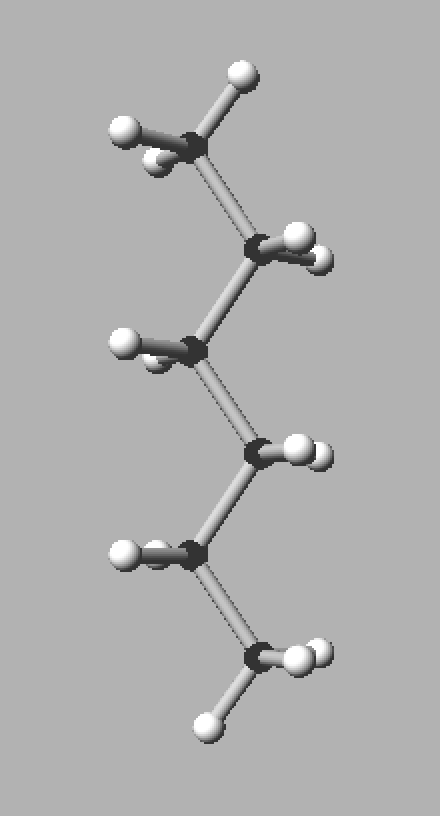}
\end {center}
\caption {Carbon chain (6 carbon atoms)}
\label {figure:carbonchain}
\end{figure}

We are considering molecules made of carbon and hydrogen atoms (linear
alkane), as shown on Fig. \ref {figure:carbonchain}. The two extremal
carbon atoms have three hydrogen atoms attached to them, while the
others have two. The number of carbon atoms is a parameter.

These molecules are coded by the
class \code {CarbonChain}. The following method builds a carbon chain with
\code {cNum} carbon atoms:

\begin{smallsize}
\begin{lstlisting}
public void build ()
   {
      buildBackbone ();
      addTop ();
      for ( int k = 1; k < cNum-1; k++ ) addH2 (k);
      addBottom ();
      createBonds ();
      createAngles ();
      createDihedrals ();
   }
\end{lstlisting}
\end{smallsize}
The backbone of carbon atoms is built by the call to
\code {buildBackbone}. Methods \code {addTop} and \code {addBottom} add 3 hydrogen atoms
to the extremities of the molecule, and \code {addH2} adds 2 hydrogens to each
carbon, except the extremities. The molecule components are created by
the 3 methods \code {createBonds}, \code {createAngles}, and
\code {createDihedrals}. The crucial point is that created molecules
have an energy which is minimal. Minimality is obtained by placing atoms at positions
compatibles with the potentials of the molecule components.
Let us consider how this is done for the two hydrogens attached to
each carbon, except the extremities. One first defines the (equilibrium)
length \code {lCH} of bonds between carbon and hydrogen atoms, and the
(equilibrium) valence angle \code {aHCH} between two hydrogens and one
carbon atoms.

\begin{smallsize}
\begin{lstlisting}
double lCH = bondC_H[1];
double aHCH = angleH_C_H[1];
double cos = lCH * Math.cos (aHCH /2);
double sin = lCH * Math.sin (aHCH /2);
\end{lstlisting}
\end{smallsize}

The \code {addH2} method is defined by:

\begin{smallsize}
\begin{lstlisting}
  void addH2 (int k)
    {
      Atom A = backbone[k-1];
      Atom B = backbone[k];
      Atom C = backbone[k+1];

      Vector3d BA = Utils.vect (B,A);
      Vector3d BC = Utils.vect (B,C);
      Vector3d P = Utils.normalize (Utils.sum (BA,BC));
      Vector3d N = Utils.normalize (Utils.perp (BA,BC));
	    
      Vector3d u = Utils.extProd (-cos,P);
      Vector3d v = Utils.extProd (-sin,N);
      Vector3d w = Utils.sum (u,v);
      Vector3d q = Utils.sum (u,Utils.opposite (v));	    	    

      Atom h1 = new H (this,Utils.sum (B.position,w),B.velocity);
      Atom h2 = new H (this,Utils.sum (B.position,q),B.velocity);	    
	    
      others [k] = new Atom [2];
      others [k][0] = h1;
      others [k][1] = h2;
  }
\end{lstlisting}
\end{smallsize}
Atoms \code {A,B,C} are three successive carbon atoms, and \code {B} is
the carbon on which two hydrogens have to be attached. 


Two hydrogens atoms \code {h1} and \code {h2} are created and placed
at their correct equilibrium positions. The two hydrogens are made accessible by
the \code {others} array of \code {B}. By this construction, the two planes
\code {h1Bh2} and \code {ABC} are orthogonal, the angle \code {h1Bh2} is
equals to \code {aHCH}, and the distances \code {h1B} and \code {h2B} are
both equal to \code {lCH}.

We now consider the creation of bonds. For the sake of simplicity,
we do not consider the other components, which are processed in a
similar manner. Bonds are created by the following method:

\begin{smallsize}
\begin{lstlisting}
void createBonds ()
   {
      for ( int k = 0; k < cNum - 1; k++ ) {
         new HarmonicBond (this,backbone[k],backbone[k+1],bondC_C);
      }
      for ( int k = 0; k < cNum; k++ ) {
          Atom c = backbone [k];
            for ( int l = 0; l < others [k].length; l++ ) {
               Atom a = others[k][l];
               if ( a instanceof H ) 
                  new HarmonicBond (this,c,a,bondC_H);
               else if ( a instanceof O ) 
                  new HarmonicBond (this,c,a,bondC_O);		    
            }
       }	
   }
\end{lstlisting}
\end{smallsize}
Lines 12-13 consider the case of oxygen atoms, to build acid
molecules, which is not considerered here.

The molecule shown on Fig. \ref {figure:carbonchain} is made of 20 atoms, 19 bonds, 36 valence
angles, and 45 dihedral angles.

\subsection {Simulations}

Reactive machines are basically provided by the class
\code{Simulation} which extends the class \code{Machine}. 
An application can simply be defined as an extension
of \code{Simulation}, as in:

\begin{smallsize}
\begin{lstlisting}
public class MinimalApp extends Simulation
{
  int cNum = 6;
  double timeStep = 1E-3;
  void molecule (double x,double y,double z)
     {
        Molecule mol = new CarbonChain (this,cNum,x,y,z,0,0,0);
	mol.context.timeStep = timeStep;
	mol.build ();
	mol.registerIn (this);
     }
  public MinimalApp ()
      {
         createUniverse ();
         double dist = 0.4;
         molecule (-dist,0.5,0);
         molecule (dist,0.5,0);
      }
   public static void main (String [] args)
      {
          standAlone (new MinimalApp ());
      }
}
\end{lstlisting}
\end{smallsize}
The number of carbon atoms \code{cNum} is set to 6 and the time-step
is set to the femto-second (lines 3 and 4; the basic time unit of the
system is the pico-second). A function which creates a
molecule is defined lines 5-11. The molecule is built and registered
in the simulation (which is denoted by \code {this}); the registration
of the molecule entails the registration of all its components. The
time-step of the created molecule is also set by the function. The
constructor of the class is defined in lines 12-18. First, the \code
{createUniverse} method provided by {\javathreed} is called to initialise the graphics, then
two molecules are created. The definition of the \code {main} Java
method terminates the definition of the class \code{MinimalApp}.

\begin{figure} [!htb]
\begin {center}
\includegraphics[height=4cm] {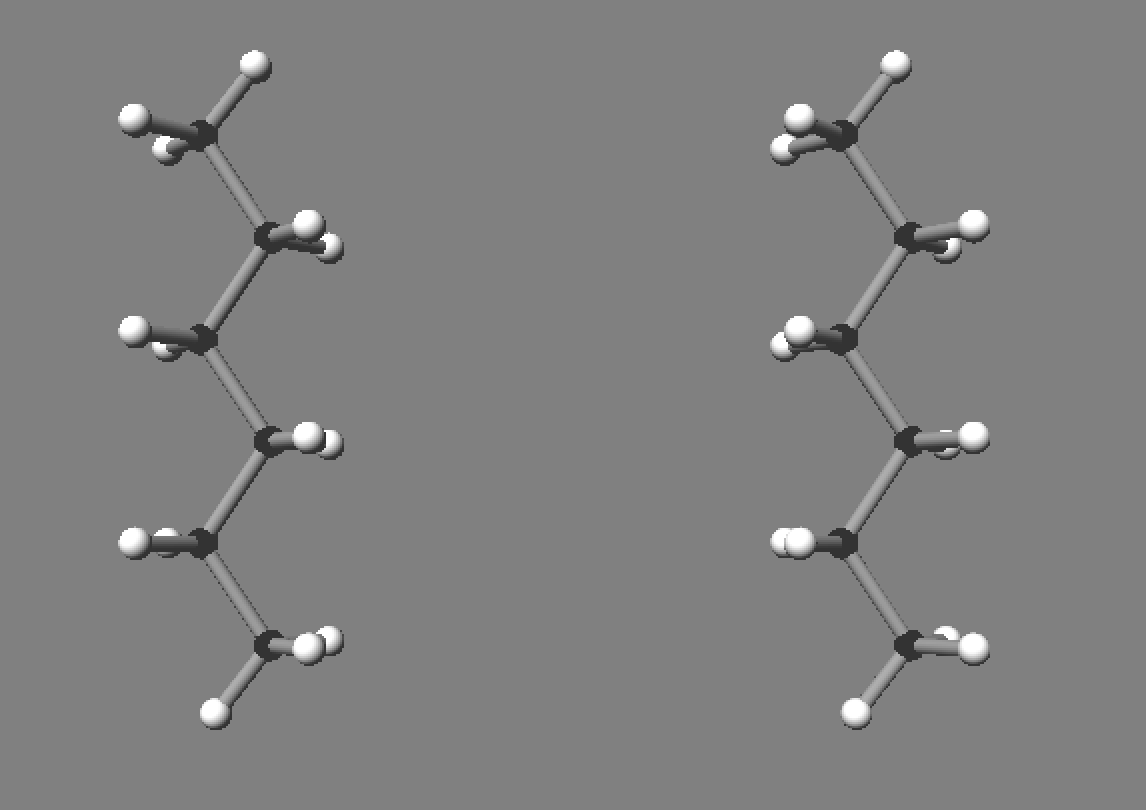} 
\includegraphics[height=4cm] {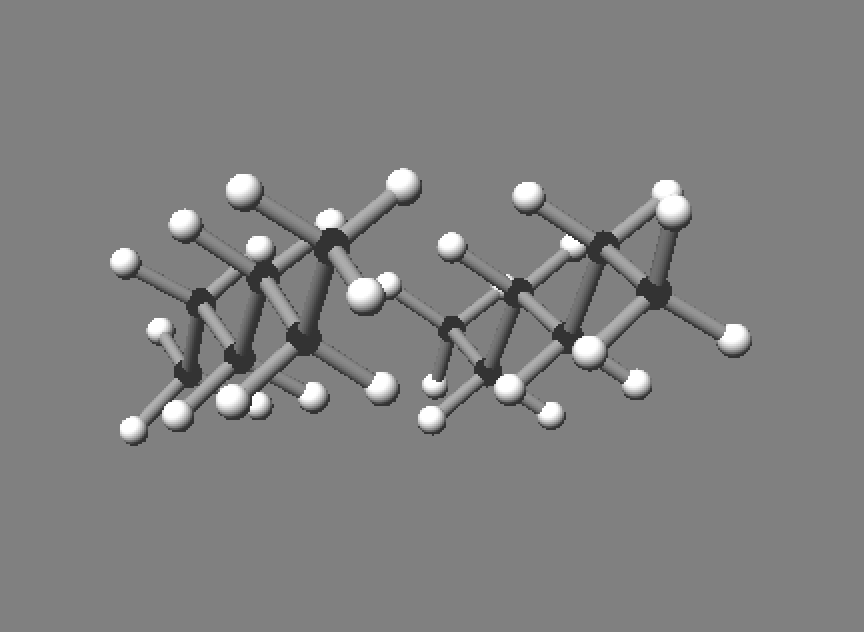}
\end {center}
\caption {Left: initial. Right: after 50 ns}
\label {figure:simul1}
\end{figure}

\begin{figure} [!htb]
\begin {center}
\includegraphics[width=11cm] {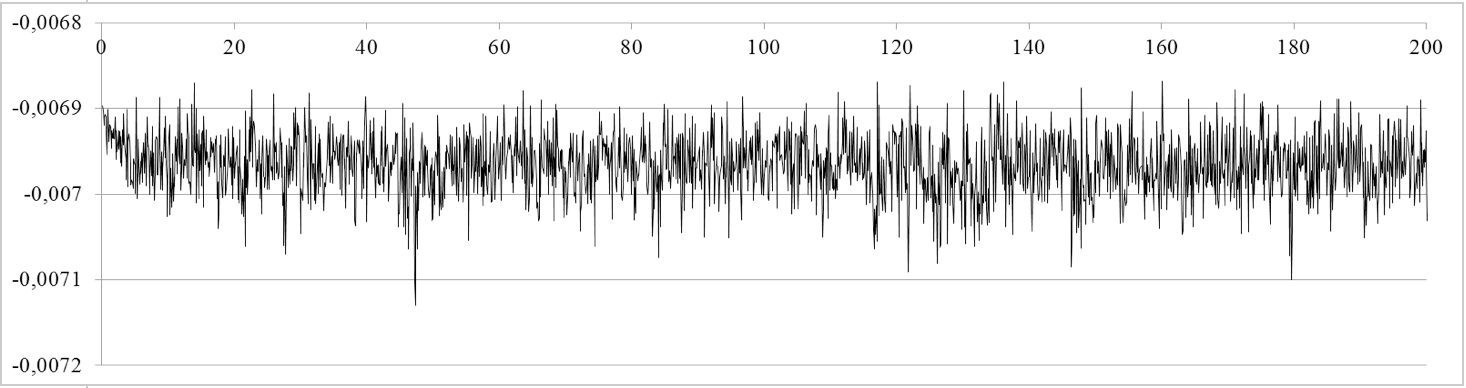}
\end {center}
\caption {Evolution of energy during 200 ns}
\label {figure:stability}
\end{figure}

The intial state of the simulation is shown on left of Fig.
\ref {figure:simul1} and the result after 50 ns ($10^8$ instants)
is shown on the right. The evolution of the energy up to 200 ns (the
internal energy unit is $kg/mol \times nm^2 / ps^2$) is shown on Fig. \ref
{figure:stability} to illustrate the stability of the resolution
(actually, stability has been tested up to one micro-second, that is
$2.10^9$ instants). The mean value is $-6.97 \times 10^{-3}$ with standard
deviation $0.035 \times 10^{-3}$. The energy is negative as result of the attraction
due to van der Waals forces.

\section {Related Work} \label {section:related-work}

The domain of physical simulations is huge and we shall thus only consider
the use of reactive programming for implementing them, and the 
implementations of MD systems.

The application of reactive programming to newtonian
physics has been initiated by Alexander Samarin in
\cite{AlexanderSamarin} where several 2D ``applets'' are proposed to
illustrate the approach.

Cellular Automata (CA) have been used in several contexts of
physics. The implementation of CA using a reactive programming
formalism is described in \cite {boussinot:inria-00071405}.

In \cite {MimickingQMwithRP} is described a system that mimicks
several aspects of quantum mechanics (namely, self-interference,
superposition of states, and entanglement). The system basically
relies on a cellular automaton plunged into a reactive based
simulation whose instants define the global time. Actually, this
cannot be strictly speaking considered as a physical simulation but
more as a kind of ``proof of concept''.

A large number of MD simulation systems exist (for example \cite
{DLPOLY} and \cite {LAMMPS}, which are both open-source
software). They are implemented in Fortran or C/C++. At the
implementation level, the focus is put on real-parallelism and the use
of multi-processor and/or multi-core architectures.  On the contrary,
we have choosen to use the Java language, and to put the focus more on
expressivity than on efficiency, by using the logical parallelism of
RP. We have adopted an open-source approach and integrated the 3D
aspects directly in the system, by using {\javathreed}.

\section {Conclusion} \label {section:conclusion}
We have shown that RP can be considered as a valuable tool for the
implementation of simulations in classical physics. We have illustated
our approach by the description of a MD system \cite {MDRP} coded in RP. We plan to
extend this MD system in several directions:

\begin {itemize}
\item Introduction of several multi-scale, multi-time-step
  aspects, building thus a true MTMS system. Note that
the dynamic creation/destruction possibilities offered by RP will be
central for the implementation of several notions (chimical reactions
and reconstruction techniques, for example).

\item Use of real-parallelism. A first study has lead to the
  definition of a new version of SugarCubes (called SugarCubesv5
  \cite{SCv5}) in which GPU-based approaches become possible. The use
  of multi-processor machines should also of course be of great
  interest.

\end {itemize}


\end{document}